\documentclass[12pt]{article}
\usepackage{epsf}
\title{ {\bf
Electric Dipole moments of charged leptons and lepton flavor violating
interactions  in the general two Higgs Doublet model}}
\author{\vspace{1cm}\\
        {\bf E. O. Iltan}
        \thanks{E-mail address:
        eiltan@heraklit.physics.metu.edu.tr}
 \\
        Physics Department, Middle East Technical University \\
        Ankara, Turkey\\}

\date{}

\begin{document}
\setlength{\baselineskip}{24pt}
\maketitle
\setlength{\baselineskip}{7mm}
\begin{abstract}
We calculate the electric dipole moment of electron using the experimental 
result of muon electric dipole moment and upper limit of the  
$BR(\mu\rightarrow e\gamma)$ in the framework of the general two Higgs 
doublet model. Our prediction is $10^{-32}\, e-cm$, which lies in the
experimental current limits. Further, we obtain constraints for the Yukawa 
couplings $\bar{\xi}^{D}_{N,\tau e}$ and $\bar{\xi}^{D}_{N,\tau\mu}$. 
Finally we present an expression which connects the 
$BR(\tau\rightarrow \mu\gamma)$ and the electric dipole moment of 
$\tau$-lepton and study the relation between these physical quantities.  
\end{abstract} 
\thispagestyle{empty}
\newpage
\setcounter{page}{1}
\section{Introduction}
Lepton Flavor Violating (LFV) interactions are one of the important source
of physics beyond the Standard model (SM) and have reached  great interest
with the improvement of experimental measurements. The processes 
$\mu\rightarrow e\gamma$ and $\tau\rightarrow \mu\gamma$ are the examples of
LFV interactions and the current limits for their branching ratios ($BR$)  
are $1.2\times 10^{-11}$ \cite{Brooks} and  $1.1\times 10^{-6}$ \cite{Ahmed}
respectively. In such decays, the assumption of 
the non-existence of Cabibbo-Kobayashi-Maskawa (CKM) type matrix in the 
leptonic sector forbids the charged Flavor Changing (FC) interactions and 
therefore the physics beyond the SM plays the main role, where the general 
two Higgs doublet model (2HDM), so called  model III, is one of the 
candidate. In this model, LFV interactions can exist at loop level with the 
help of the internal neutral higgs bosons $h_0$ and $A_0$. The Yukawa 
couplings appearing in these loops are free parameters and their strength 
can be detemined by the experimental data. In the literature, there are 
several studies on LFV interactions in different models. Such interactions 
are studied in a model independent way in \cite{Chang}, in the framework of 
model III 2HDM \cite{Diaz}, in supersymmetric models \cite{Barbieri1,
Barbieri2,Barbieri3,Ciafaloni,Duong,Couture,Okada}. 

CP violating effects also provide comprehensive information in the  
determination of free parameters of the various theoretical models.   
In extensions of the SM, new source of CP-violating phases occur. The 
exchanges  of neutral Higgs bosons in the 2HDM \cite{Branco} or 
the charged sector in multi Higgs doublet models (more than 
two Higgs doublets) \cite{Weinberg1} induce CP violating effects. 
In the model III 2HDM, one of the source for the CP violation is the 
complex Yukawa coupling. Non-zero Electric Dipole Moments (EDM) of the 
elementary particles are the sign of such violation. These are interesting 
from the experimental point of view since there are improvements in the 
experimental limits of charged lepton EDM. EDM of electron, muon and tau 
have been measured and the present limits are  
$d_e =1.8\pm 1.2\pm 1.0 \times 10^{-27}$ \cite{Commins}, 
$d_{\mu} =(3.7\pm 3.4)\times 10^{-19}$ \cite{Bailey} and 
$d_{\tau} =3.1\times10^{-16}$ \cite{Groom}. 

Dipole moments of leptons have been studied in the literature extensively. 
In \cite{Barr}, it is emphasized that the dominant contribution to the EDM 
of lighter leptons comes from two loop diagrams that involve one power of 
the Higgs Yukawa couplings. In this work, the CP-violation is assumed to be 
mediated by neutral Higgs scalars \cite {Weinberg2} and the EDM of electron 
is predicted at the order of the magnitude of $10^{-26}\,e-cm$. 
Further EDM of leptons have been analyzed in supersymmetric 
models \cite{Barbieri2,Okada,Babu2}. In the recent work \cite{Babu} EDM 
of leptons are studied by scaling them with corresponding lepton masses 
and the EDM of electron is predicted as $10^{-27}\, e-cm$.

In our work, we study EDM of leptons $e,\mu,\tau$ and LFV processes 
$\mu\rightarrow e\gamma$, $\tau\rightarrow\mu\gamma$ in the general 2HDM 
(model III). The source of EDM of a particle is the CP violated interaction 
and it can come from the complex Yukawa couplings. In the model III, 
it is possible to get a considerable EDM at one-loop level. Further, LFV 
processes can exist also at one-loop level with internal mediating neutral
particles $h_0$ and $A_0$ since there is no CKM type matrix and therefore 
no charged FC interaction in the leptonic sector according to our assumption.

The paper is organized as follows:
In Section 2, we present EDMs of $e,\mu,\tau$ leptons and  the expressions
for the decay width of processes $\mu\rightarrow e\gamma$ and 
$\tau\rightarrow \mu\gamma$ in the framework of the model III. Section 3 is 
devoted to discussion and our conclusions.
\section{Electric dipole moments of charged leptons and LFV interactions 
in the general two Higgs Doublet model.} 
The 2HDM type III permits flavor changing neutral currents (FCNC) at tree 
level and makes CP violating interactions possible with the choice of complex 
Yukawa couplings. The  Yukawa interaction for the leptonic sector in the model III is
\begin{eqnarray}
{\cal{L}}_{Y}=
\eta^{D}_{ij} \bar{l}_{i L} \phi_{1} E_{j R}+
\xi^{D}_{ij} \bar{l}_{i L} \phi_{2} E_{j R} + h.c. \,\,\, ,
\label{lagrangian}
\end{eqnarray}
where $i,j$ are family indices of leptons, $L$ and $R$ denote chiral 
projections $L(R)=1/2(1\mp \gamma_5)$, $\phi_{i}$ for $i=1,2$, are the 
two scalar doublets, $l_{i L}$ and $E_{j R}$ are lepton doublets and
singlets respectively. 
Here $\phi_{1}$ and $\phi_{2}$ are chosen as
\begin{eqnarray}
\phi_{1}=\frac{1}{\sqrt{2}}\left[\left(\begin{array}{c c} 
0\\v+H^{0}\end{array}\right)\; + \left(\begin{array}{c c} 
\sqrt{2} \chi^{+}\\ i \chi^{0}\end{array}\right) \right]\, ; 
\phi_{2}=\frac{1}{\sqrt{2}}\left(\begin{array}{c c} 
\sqrt{2} H^{+}\\ H_1+i H_2 \end{array}\right) \,\, ,
\label{choice}
\end{eqnarray}
and the vacuum expectation values are  
\begin{eqnarray}
<\phi_{1}>=\frac{1}{\sqrt{2}}\left(\begin{array}{c c} 
0\\v\end{array}\right) \,  \, ; 
<\phi_{2}>=0 \,\, .
\label{choice2}
\end{eqnarray}
With this choice, the SM particles can be collected in the first doublet 
and the new particles in the second one. The part which produce FCNC at tree
level is  
\begin{eqnarray}
{\cal{L}}_{Y,FC}=
\xi^{D}_{ij} \bar{l}_{i L} \phi_{2} E_{j R} + h.c. \,\, .
\label{lagrangianFC}
\end{eqnarray}
Here the Yukawa matrices $\xi^{D}_{ij}$ have in general complex entries. 
Note that in the following we replace $\xi^{D}$ with $\xi^{D}_{N}$ where 
"N" denotes the word "neutral". The complex Yukawa couplings are the source 
of CP violation and EDM of particles are created by the CP violated 
interactions.  The effective EDM interaction for a lepton is defined as 
\begin{eqnarray}
{\cal L}_{EDM}=i d_l \,\bar{l}\,\gamma_5 \,\sigma_{\mu\nu}\,l\, F^{\mu\nu} 
\,\, ,
\label{EDM1}  
\end{eqnarray}
where $F_{\mu\nu}$ is the electromagnetic field tensor and "$d_l$" is EDM 
of the lepton. Here, "$d_l$" is a real number by hermiticity. Concentrating
only on neutral currents in the model III, the neutral Higgs bosons $h_0, 
A_0$ can induce CP violating interactions which can create EDM at loop
level. Note that, we take $H_1$ and $H_{2}$ in eq. (\ref{choice}) as the 
mass eigenstates $h_0$ and $A_0$ respectively since no mixing between
CP-even neutral Higgs bosons $h_0$ and the SM one, $H_0$, occurs at tree 
level. Due to possible small mixing at loop level we also study the mixing
effects on the physical parameters in the Discussion part. 

\begin{figure}[htb]
\vskip -3.0truein
\centering
\epsfxsize=6.8in
\leavevmode\epsffile{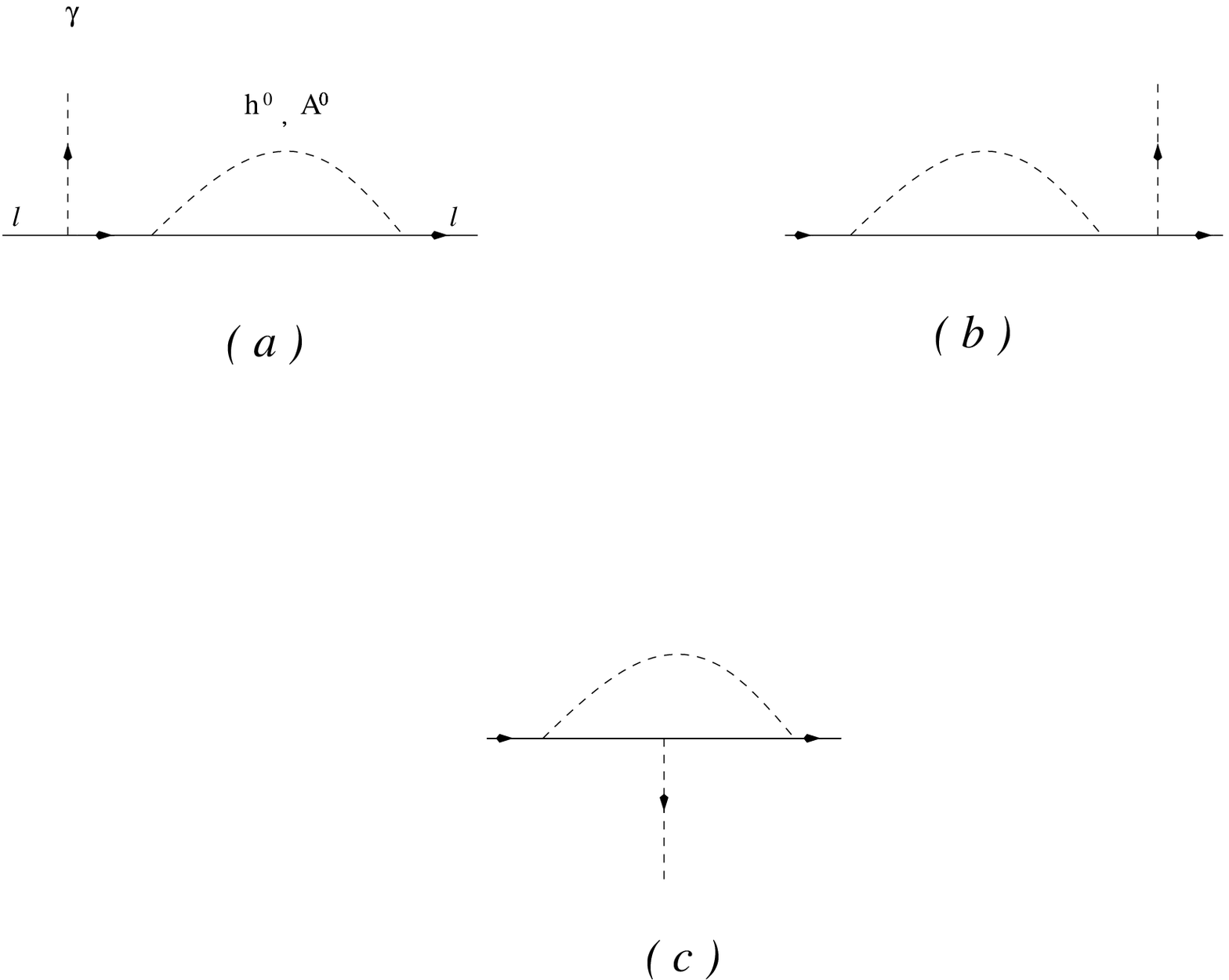}
\vskip -7.0truein
\caption[]{One loop diagrams contribute to EDM of $l$-lepton and LFV
interactions (if external leptons $l$ have different flavors) due to the 
neutral Higgs bosons $h_0$ and $A_0$ in the 2HDM. Dashed lines represent 
the electromagnetic field, $h_0$ and $A_0$ fields.}
\label{fig1}
\end{figure}

Now, we give the necessary 1-loop diagrams due to neutral Higgs 
particles in Fig. \ref{fig1}.  Since, in the on-shell renormalization scheme, 
the self energy $\sum(p)$ can be written as 
\begin{eqnarray}
\sum(p)=(\hat{p}-m_l)\bar{\sum}(p) (\hat{p}-m_l)\,\, ,
\label{self}
\end{eqnarray}
diagrams $a$, $b$ in Fig. \ref{fig1} vanish when $l$-lepton is on-shell. 
However the vertex diagram $c$ in Fig. \ref{fig1} gives non-zero 
contribution. The most general Lorentz-invariant form of the coupling of 
a charged lepton to a photon of 4-momentum $q_{\nu}$ can be written as
\begin{eqnarray}
\Gamma_{\mu}&=& G_1(q^2)\, \gamma_{\mu}+
+ G_2 (q^2)\, \sigma_{\mu\nu} \,q^{\nu} \nonumber \\
&+& G_3 (q^2)\, 
\sigma_{\mu\nu}\gamma_5\, q^{\nu}
\label{vertexop}
\end{eqnarray}
where $q_{\nu}$ is photon 4-vector and $q^2$ dependent form factors 
$G_{1}(q^2)$ and  $G_{2}(q^2)$ are proportional to the charge and 
anamolous magnetic moment of $l$-lepton respectively. Non-zero value of 
$G_{3}(q^2)$ is responsible for the CP violation and it is proportional 
to EDM of $l$-lepton. By extracting this  part of the vertex, $l$-lepton 
EDM "$d_l$" $(l=e,\,\mu,\,\tau)$ (see eq.(\ref{EDM1})) can be calculated 
as a sum of contributions coming from neutral Higgs bosons $h_0$ and $A_0$,
\begin{eqnarray}
d_l= -\frac{i G_F}{\sqrt{2}} \frac{e}{32\pi^2}\, 
\frac{Q_{\tau}}{m_{\tau}}\, ((\bar{\xi}^{D\,*}_{N,l\tau})^2-
(\bar{\xi}^{D}_{N,\tau l})^2)\,
(F_1 (y_{h_0})-F_1 (y_{A_0})) \,\, ,
\label{emuEDM}
\end{eqnarray}
for $l=e,\mu$ and 
\begin{eqnarray}
d_{\tau}&=&
-\frac{i G_F}{\sqrt{2}} \frac{e}{32\pi^2}\, \Big{\{} 
\frac{Q_{\tau}}{m_{\tau}}\, ((\bar{\xi}^{D\,*}_{N,\tau\tau})^2-
(\bar{\xi}^{D}_{N,\tau \tau})^2)\, 
(F_2 (r_{h_0})-F_2(r_{A_0}))\nonumber \\ 
&-&
Q_{\mu}\, \frac{m_{\mu}}{m^2_{\tau}} 
((\bar{\xi}^{D\,*}_{N,\mu\tau})^2-(\bar{\xi}^{D}_{N,\tau \mu})^2)\,
(r_{h_0}\,ln\, (z_{h_0})-r_{A_0}\,ln\, (z_{A_0})) \Big{\}}
\,\, ,
\label{tauEDM}
\end{eqnarray}
where the functions $F_1 (w)$, $F_2 (w)$ and $F_3 (w)$
are 
\begin{eqnarray}
F_1 (w)&=&\frac{w\,(3-4\,w+w^2+2\,ln\,w)}{(-1+w)^3}\nonumber \,\, , \\
F_2 (w)&=& w\, ln\,w + \frac{2\,(-2+w)\, w\,ln\,
\frac{1}{2}(\sqrt{w}-\sqrt{w-4})}{\sqrt{w\,(w-4)}} \,\, .
\label{functions1}
\end{eqnarray}
Here  $y_{H}=\frac{m^2_{\tau}}{m^2_{H}}$, $r_{H}=\frac{1}{y_{H}}$ and 
$z_{H}=\frac{m^2_{\mu}}{m^2_{H}}$, $Q_{\tau}$ and $Q_{\mu}$ are charges 
of $\tau$ and $\mu$ leptons respectively. In eqs. (\ref{emuEDM}) and 
(\ref{tauEDM}) $\bar{\xi}^{D}_{N,ij}$ is defined as 
$\xi^{D}_{N,ij}=\sqrt{\frac{4\,G_F}{\sqrt{2}}}\, \bar{\xi}^{D}_{N,ij}$.
In eq. (\ref{emuEDM}) we take into account only internal $\tau$-lepton 
contribution since, in our assumption, the Yukawa couplings 
$\bar{\xi}^{D}_{N, ij},\, i,j=e,\mu$, are small compared to 
$\bar{\xi}^{D}_{N,\tau\, i}\, i=e,\mu,\tau$ due to the possible
proportionality of the Yukawa couplings to the masses of leptons
underconsideration in the vertices. In eq. (\ref{tauEDM}) we take also 
the internal $\mu$-lepton contribution besides internal $\tau$-lepton one 
and ignore the one coming from the internal $e$-lepton respecting our
assumption (see Discussion part). Note that, we make our calculations in
arbitrary $q^2$ and take $q^2=0$ at the end.   

Using the parametrization 
\begin{eqnarray}
\bar{\xi}^{D}_{N,\tau l}=|\bar{\xi}^{D}_{N,\tau l}|\, e^{i \theta_{l}}
\,\, , 
\label{xi}
\end{eqnarray}
the Yukawa factors in eqs. (\ref{emuEDM}) and  (\ref{tauEDM}) can be 
written as 
\begin{eqnarray}
((\bar{\xi}^{D\,*}_{N,l\tau})^2-(\bar{\xi}^{D}_{N,\tau l})^2)=-2\,i
sin\,2\theta_{l}\, |\bar{\xi}^{D}_{N,\tau l}|^2
\end{eqnarray}
where $l=e,\mu,\tau$. Here $\theta_{l }$ are CP violating parameters 
which are the sources of the lepton EDM. 

Now, let us consider the lepton number violating process 
$\mu\rightarrow e\gamma$ which is a good candidate in the determination of
the Yukawa couplings and new physics beyond the SM. Since we take into
account only the neutral Higgs contributions in the lepton sector of the 
model III, the contribution comes from neutral Higgs bosons $h_0$ and 
$A_0$ (see Fig. \ref{fig1}). In the on-shell renormalization scheme the 
self energy diagrams are cancelled and the vertex diagram 
(Fig. \ref{fig1}-c) contributes. Taking only $\tau$ lepton for the internal 
line, the decay width $\Gamma$ reads as 
\begin{eqnarray}
\Gamma (\mu\rightarrow e\gamma)=c_1(|A_1|^2+|A_2|^2)\,\, ,
\label{DWmuegam}
\end{eqnarray}
where
\begin{eqnarray}
A_1&=&Q_{\tau} \frac{1}{8\,m_{\mu}\,m_{\tau}} \bar{\xi}^D_{N,\tau e}\, 
\bar{\xi}^D_{N,\tau \mu}\, (F_1 (y_{h_0})-F_1 (y_{A_0}))
\nonumber \,\, , \\
A_2&=&Q_{\tau} \frac{1}{8\,m_{\mu}\,m_{\tau}} \bar{\xi}^{D *}_{N,e \tau}\, 
\bar{\xi}^{D *}_{N,\mu \tau}\, (F_1 (y_{h_0})-F_1 (y_{A_0})) \,\, ,
\label{A1A2}
\end{eqnarray}
$c_1=\frac{G_F^2 \alpha_{em} m^5_{\mu}}{32 \pi^4}$ and the function 
$F_1 (w)$ is given in eq. (\ref{functions1}). Here the amplitudes $A_1$ and
$A_2$ have right and left chirality  respectively. In eq. (\ref{DWmuegam}) 
we ignore the contributions coming from internal $\mu$ and $e$ leptons 
respecting our assumption on the Yukawa couplings (see Discussion).  

Another LFV process is $\tau\rightarrow \mu\gamma$ and it is 
rich from the theoretical point of view. The decay width of this process can 
be calculated using the same procedure and reads as 
\begin{eqnarray}
\Gamma (\tau\rightarrow \mu\gamma)=c_2(|B_1|^2+|B_2|^2)\,\, ,
\end{eqnarray}
\label{DWtaumugam}
where
\begin{eqnarray}
B_1&=& Q_{\tau} \frac{1}{48\,m_{\mu}\,m_{\tau}} \bar{\xi}^D_{N,\tau \mu} 
\Big{\{}\bar{\xi}^{D *}_{N,\tau \tau} (G_1 (y_{h_0})+G_1 (y_{A_0}))+ 
6 \bar{\xi}^{D}_{N,\tau \tau} (F_1 (y_{h_0})-F_1 (y_{A_0})\Big{\}} 
\nonumber \,\, , \\
B_2&=& Q_{\tau} \frac{1}{48\,m_{\mu}\,m_{\tau}} \bar{\xi}^{D *}_{N,\mu\tau} 
\Big{\{}\bar{\xi}^{D}_{N,\tau \tau} (G_1 (y_{h_0})+G_1 (y_{A_0}))+ 
6 \bar{\xi}^{D *}_{N,\tau \tau} (F_1 (y_{h_0})-F_1 (y_{A_0})\Big{\}} 
\,\, ,
\label{B1B2}
\end{eqnarray}
and $c_2=\frac{G_F^2 \alpha_{em} m^5_{\tau}}{32 \pi^4}$. Here the amplitudes 
$B_1$ and $B_2$ have right and left chirality  respectively. The function 
$G_1 (w)$ is given by 
\begin{eqnarray}
G_1 (w)&=&\frac{w\,(2+3\,w-6\,w^2+w^3+ 6\,w\,ln\,w)}{(-1+w)^4}
\nonumber \,\, . 
\label{functions2}
\end{eqnarray}
Note that, in the case of the possible loop induced mixing of the neutral
Higgs bosons $h_0$ and the SM one $H_0$, the functions $F_1 (y_{h_0})$, 
$F_2 (r _{h_0})$, $ln (z_{h_0})$ and $G_1 (y_{h_0})$ in eqs. 
(\ref{emuEDM}, \ref{tauEDM}, \ref{A1A2}, \ref{B1B2}) should be multiplied 
by $cos^2\,\alpha$ where $\alpha$ is the small mixing angle. 
%
\section{Discussion}
The Yukawa couplings $\bar{\xi}^D_{N,ij}, i,j=e, \mu, \tau$ play the main 
role in the determination of the lepton EDM and the physical quantities for 
LFV interactions, in the model III. These couplings are complex in general
and they are free parameters which can be fixed by present and forthcoming 
experiments. The experimental results on the EDM of leptons $e,\mu, \tau$ 
and LFV process $\mu\rightarrow e\gamma$ are our starting point of the 
predictions on Yukawa couplings related with leptons. Note that, in our 
predictons, we assume that the Yukawa couplings 
$\bar{\xi}^{D}_{N,ij},\, i,j=e,\mu $, are small compared to 
$\bar{\xi}^{D}_{N,\tau\, i}\, i=e,\mu,\tau$ since the strength of these
couplings are related with the masses of leptons denoted by the indices of 
them, similar to the Cheng-Sher scenerio \cite{Sher}. Further, we assume 
that $\bar{\xi}^{D}_{N,ij}$ is symmetric with respect to the indices $i$ 
and $j$.   

\begin{figure}[htb]
\vskip -3.0truein
\centering
\epsfxsize=6.8in
\leavevmode\epsffile{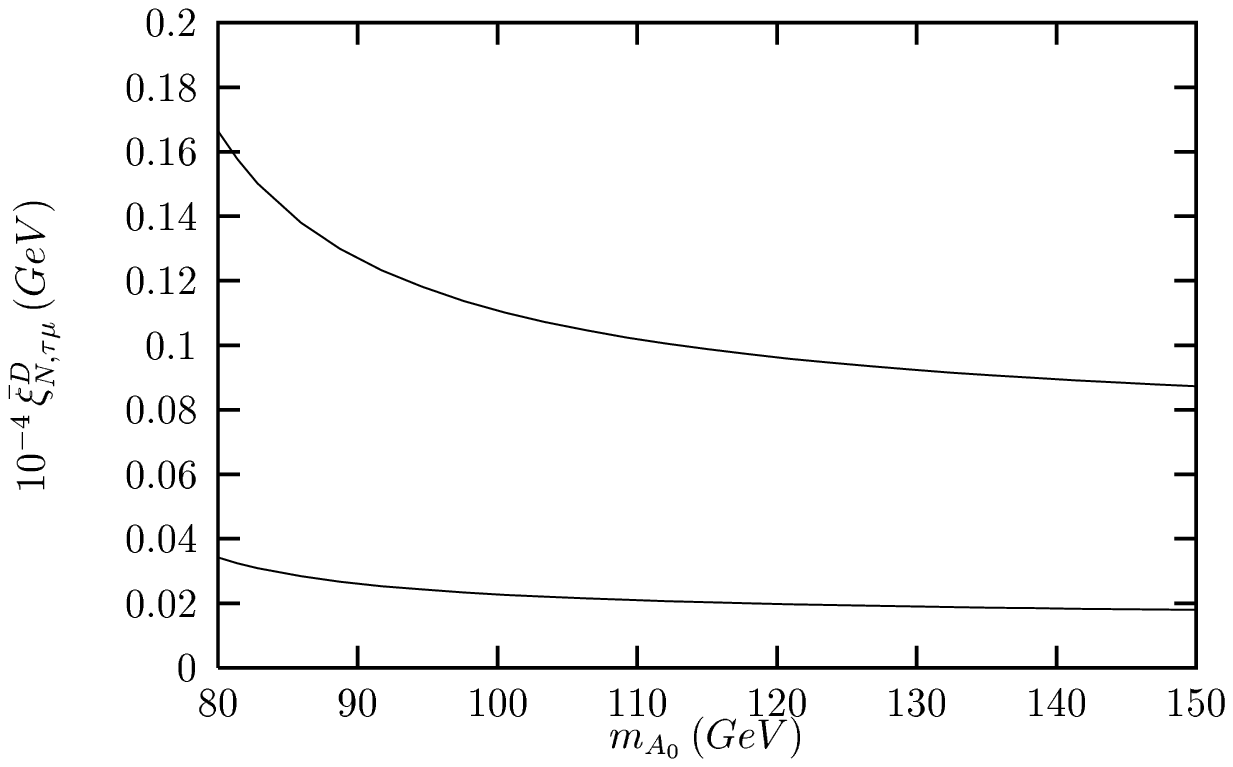}
\vskip -3.0truein
\caption[]{$\bar{\xi}_{N,\tau\mu}^{D}$ as a function of $m_{A_0}$ 
for $m_{h_0}=70\, GeV$ and $sin\,\theta_{\mu} = 0.5$.
Here the coupling is restricted in the region between solid lines.} 
\label{ksiDtaumuA0}
\end{figure}

Since non-zero EDM can be obtained in case of complex 
couplings, there exist a CP violating parameter $\theta_l$ coming from 
the parametrization eq. (\ref{xi}). At this stage, we find a restriction 
for $\bar{\xi}^{D}_{N,\tau\mu}$ using eq. (\ref{emuEDM}) and the experimental 
result of $\mu$ EDM \cite{Abdullah},
\begin{eqnarray}
0.3\times 10^{-19}\, e-cm < d_{\mu} < 7.1\times 10^{-19}\, e-cm \,\, .
\label{muedmex}
\end{eqnarray}
Fig. \ref{ksiDtaumuA0} shows $m_{A_0}$ dependence of 
$\bar{\xi}^{D}_{N,\tau\mu}$ for $sin\theta_{\mu}=0.5$ and $m_{h_0}=70\, GeV$. 
Here the coupling is restricted in the region between solid lines. As shown 
in the figure, $\bar{\xi}^{D}_{N,\tau\mu}$ is at the order of the magnitude 
$10^{2}$ and take smaller values if $m_{A_0}$ takes larger values compared 
to $m_{h_0}$ in the theory. If neutral Higgs masses are almost degenerate, 
namely $m_{A_0}=m_{h_0}$, $\bar{\xi}^{D}_{N,\tau\mu}$ should be very large 
since the contributions of $h_0$ and $A_0$ have opposite signs and the same 
functional dependences. Therefore to be able control the numerical value of 
the coupling $\bar{\xi}^{D}_{N,\tau\mu}$, the masses should not be degenerate.  

\begin{figure}[htb]
\vskip -3.0truein
\centering
\epsfxsize=6.8in
\leavevmode\epsffile{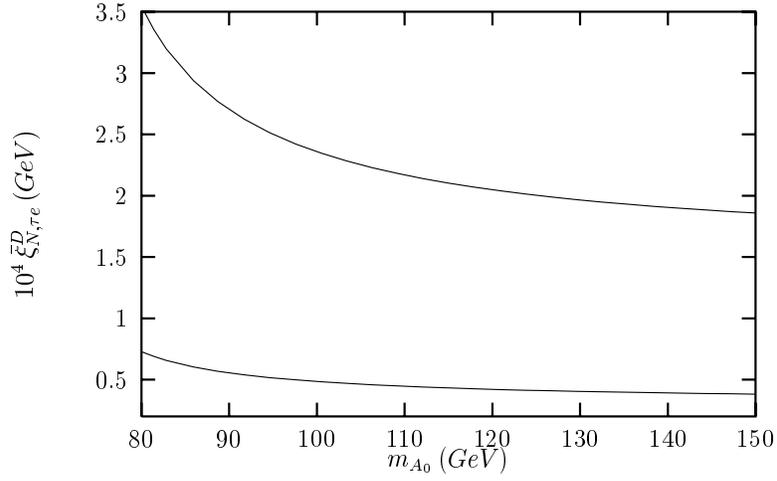}
\vskip -3.0truein
\caption[]{The same as Fig. \ref{ksiDtaumuA0} but for 
$\bar{\xi}_{N,\tau e}^{D}$.}
\label{ksiDtaueA0}
\end{figure}

\begin{figure}[htb]
\vskip -3.0truein
\centering
\epsfxsize=6.8in
\leavevmode\epsffile{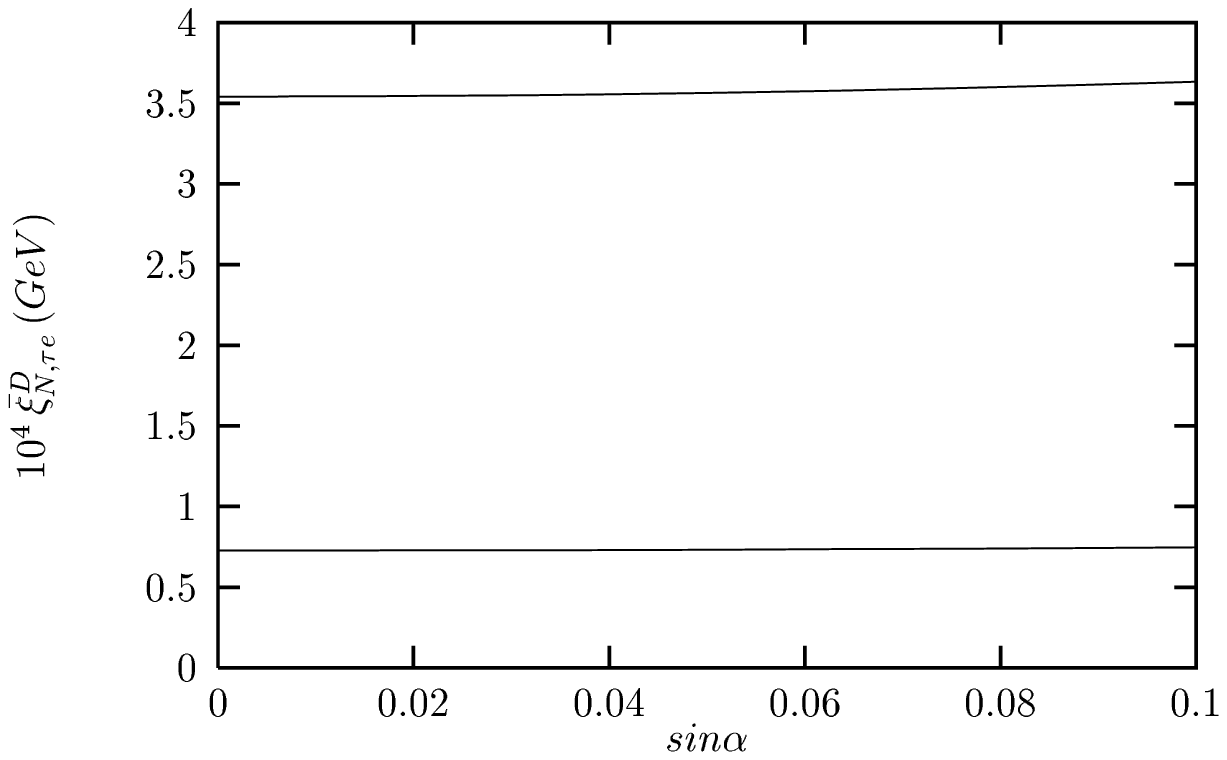}
\vskip -3.0truein
\caption[]{$\bar{\xi}_{N,\tau e}^{D}$ as a function of $sin\,\alpha$ 
for $m_{h_0}=70\, GeV$, $m_{A_0}=80\, GeV$ and $sin\,\theta_{\mu} = 0.5$.
Here the coupling is restricted in the region between solid lines.}
\label{ksiDtauealf}
\end{figure}

Now, we use the upper limit of the $BR$ of the process 
$\mu\rightarrow e\gamma$ to restrict the Yukawa combination   
$\bar{\xi}^{D}_{N,\tau\mu}\,\bar{\xi}^{D}_{N,\tau e}$ (see eq. 
(\ref{A1A2})). Here we take into account only the internal $\tau$-lepton 
contribution. Using the restrictions for $\bar{\xi}^{D}_{N,\tau\mu}$ and 
$\bar{\xi}^{D}_{N,\tau\mu}\,\bar{\xi}^{D}_{N,\tau e}$ we find a constraint 
region for $\bar{\xi}^{D}_{N,\tau e}$. In Fig. \ref{ksiDtaueA0} we plot 
$\bar{\xi}^{D}_{N,\tau e}$ as a function of $m_{A_0}$ for 
$sin\theta_{\mu}=0.5$ and $m_{h_0}=70\, GeV$. Here the coupling is 
restricted in the region between solid lines. As shown in the figure 
$\bar{\xi}^{D}_{N,\tau e}$ is at the order of the magnitude $10^{-4}$ and 
becomes smaller if the mass of the particle $A_0$ is larger compared 
to $m_{h_0}$ in the theory, similar to the coupling 
$\bar{\xi}^{D}_{N,\mu\tau}$. Fig. \ref{ksiDtauealf} represente the effect of
possible mixing between CP even neutral Higgs bosons $h_0$ and $H_0$ on 
the coupling $\bar{\xi}^{D}_{N,\tau e}$. When the mixing parameter 
$sin\,\alpha$ changes in the range $0-0.1$, the upper limit of  
$\bar{\xi}^{D}_{N,\tau e}$ is affected at the order of the magnitude $3\%$ 
for the fixed values of $sin\theta_{\mu}=0.5$ , $m_{h_0}=70\, GeV$ and 
$m_{A_0}=80\, GeV$. The lower limit of the coupling is almost non-sensitive
to the mixing as shown in the figure.   

\begin{figure}[htb]
\vskip -3.0truein
\centering
\epsfxsize=6.8in
\leavevmode\epsffile{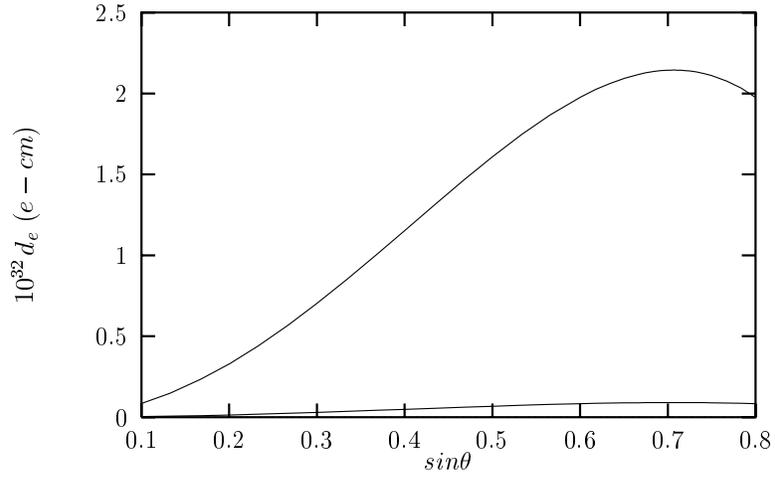}
\vskip -3.0truein
\caption[]{$d_e$ as a function of  $sin\,\theta_{e}$ for 
$m_{h_0}=70\, GeV$ and  $m_{A_0}=80\, GeV$. 
Here $d_e$ is restricted in the region bounded by solid lines.} 
\label{EDMesin}
\end{figure}

\begin{figure}[htb]
\vskip -3.0truein
\centering
\epsfxsize=6.8in
\leavevmode\epsffile{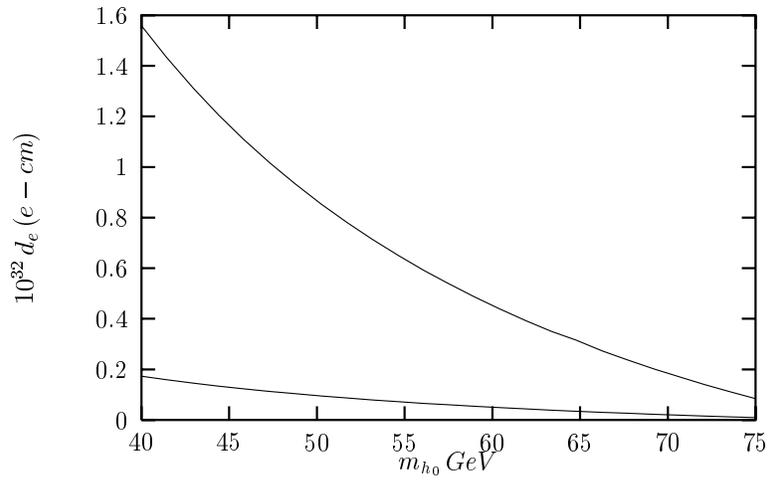}
\vskip -3.0truein
\caption[]{The same as Fig. \ref{EDMesin} but $d_e$ as a function of  
$m_{h_0}$ for $sin\,\theta_{e}=0.5$. } 
\label{EDMeh0}
\end{figure}

\begin{figure}[htb]
\vskip -3.0truein
\centering
\epsfxsize=6.8in
\leavevmode\epsffile{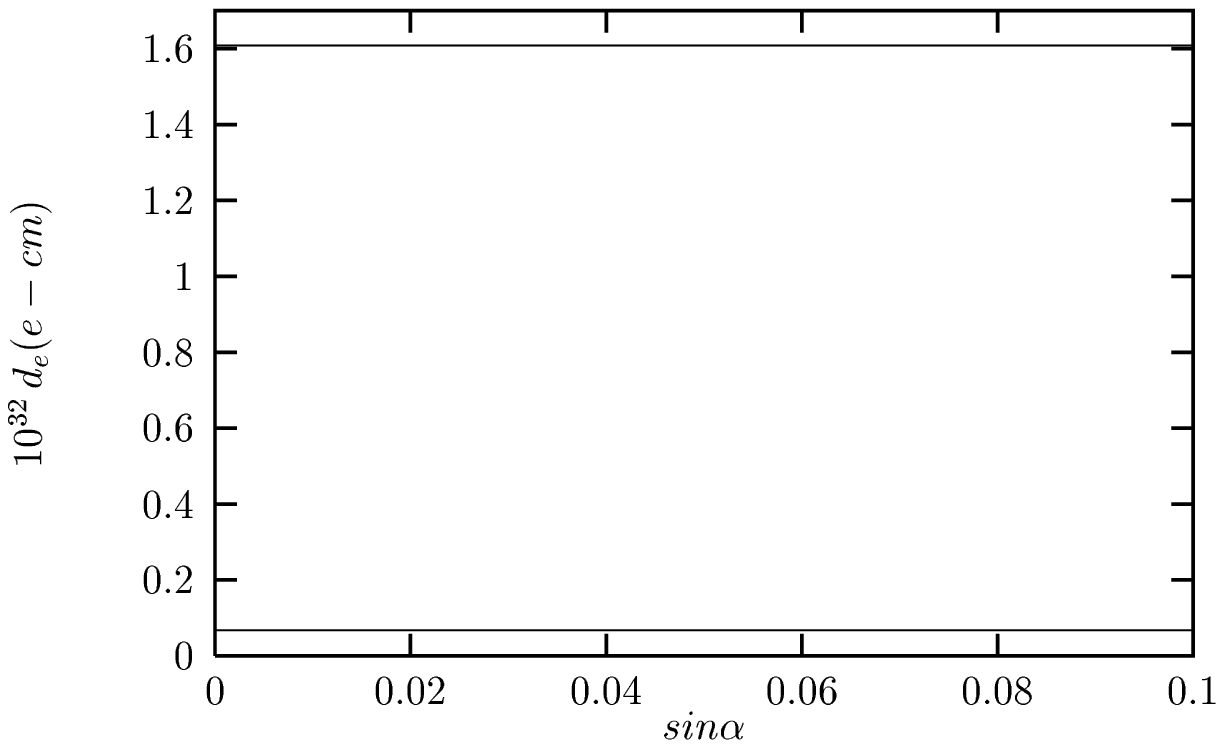}
\vskip -3.0truein
\caption[]{$d_e$ as a function of  $sin\,\alpha$ for 
$m_{h_0}=70\, GeV$, $m_{A_0}=80\, GeV$ and $sin\,\theta_{e}=0.5$. 
Here $d_e$ is restricted in the region bounded by solid lines.} 
\label{EDMealf}
\end{figure}

At this stage we are ready to predict the electron EDM, $d_e$. Using eq. 
(\ref{emuEDM}) and the restriction for $\bar{\xi}^{D}_{N,\tau e}$, $d_e$ is 
plotted with respect to $\sin\theta_{e}$ in Fig. \ref{EDMesin} for fixed 
values of $sin\theta_{\mu}=0.5$, $m_{h_0}=70\, GeV$ and $m_{A_0}=80\, GeV$. 
For the intermediate values of $sin\theta_{e}$, $d_e$ is at the order of the 
magnitude $10^{-32}$ which lies in the experimental restriction region 
\cite{Commins} 
\begin{eqnarray}
d_{e} =1.8\pm 1.2\pm 1.0 \times 10^{-27}\, e-cm
\label{eedmex}
\end{eqnarray}
We also present $m_{h_0}$ dependence of $d_{e}$ for fixed values of 
$sin\theta_{e}=0.5$ and $m_{A_0}=80\,GeV$ in Fig. \ref{EDMeh0}. This figure 
shows that $d_{e}$ is bounded by the solid lines and decreases with 
increasing values of $m_{h_0}$. For completeness, we study $h_0$-$H_0$ 
mixing effect on $d_{e}$. Fig. \ref{EDMealf} shows that $d_{e}$ is not
sensitive to mixing of two CP-even bosons in the given range of $
sin\,\alpha$. 

Finally, we would like to predict the behaviour of $d_{\tau}$ with respect
to $BR(\tau\rightarrow \mu\gamma)$. Since we take into account internal 
$\tau$ and $\mu$ leptons in the calculation of $d_{\tau}$, it has a 
functional dependence (see eq. (\ref{tauEDM}))
\begin{eqnarray}
d_{\tau}=f_1+ f_2 |\bar{\xi}^{D}_{N,\tau\tau}|^2 \, \, ,
\label{dtau2}
\end{eqnarray}
where $f_1$ contains the coupling $\bar{\xi}^{D}_{N,\tau\mu}$ and its
complex conjugate, which can be fixed by the constraint plotted in Fig. 
\ref{ksiDtaumuA0}. On the other hand $BR (\tau\rightarrow\mu\gamma)$
has in the form
\begin{eqnarray}
BR(\tau\rightarrow\mu\gamma)=g_1 |\bar{\xi}^{D}_{N,\tau\mu}|^2 \, 
|\bar{\xi}^{D}_{N,\tau\tau}|^2 \,\, ,
\label{BRtau2}
\end{eqnarray}
where $|\bar{\xi}^{D}_{N,\tau\mu}|^2$ is again fixed by the constraint 
we have. As a result, we connect $d_{\tau}$ and \\
$BR(\tau\rightarrow\mu\gamma)$ by an expression in the form 
\begin{eqnarray}
d_{\tau}=f_1+ g_2 BR(\tau\rightarrow\mu\gamma)\,\, .
\label{dtauBR}
\end{eqnarray}
Here the lower limit of $f_1$ is at the order of the magnitude of 
$10^{-15}$ for the fixed values of $m_{h_0}=70\, GeV$, $m_{A_0}=80\, GeV$ 
and $sin\theta_{\mu}=0.5$. Therefore, even internal $\mu$-lepton causes to
exceed the upper limit of experimental value of $d_{\tau}$ \cite{Groom} 
\begin{eqnarray}
d_{\tau} = 3.1\times 10^{-16}\, e-cm \,\, ,
\label{tauedmex}
\end{eqnarray}
In eq. (\ref{dtauBR}), the part $g_2\, BR(\tau\rightarrow\mu\gamma)$ is due 
to the internal $\tau$-lepton contribution. Here $g_2$ is at the order of 
the magnitude of $10^{-14}$. Therefore it is not possible to get a strict
bound for $BR(\tau\rightarrow\mu\gamma)$ using the eqs. (\ref{dtauBR}) and
the experimental result eq. (\ref{tauedmex}).

In our work we choose the 2HDM type III and assume that only FCNC 
interactions exist at tree level. First we obtain constraints for the 
couplings $\bar{\xi}^{D}_{N,\tau e}$ and $\bar{\xi}^{D}_{N, \tau\mu}$. 
The previous one is at the order of the magnitude of $10^{-4}$ and the 
last one is $10^{2}$. Numerical values of these couplings decrease when
the mass difference of neutral Higgs bosons $h_0$ and $A_0$ increases. 
Further, we predict the $d_e$ as $10^{-32}\, e-cm$ using 
the experimental result of $d_{\mu}$ and the upper limit of 
$BR(\mu\rightarrow e\gamma)$. This lies in the experimental restriction 
region of $d_e$, however it is far from the present limits.  Finally, we 
calculated the $BR(\tau\rightarrow \mu\gamma)$ and $d_{\tau}$ and connect 
these physical quantities in the form of a linear expression 
(eq. (\ref{dtauBR})). This expression does not provide a strict bound for 
$BR(\tau\rightarrow \mu\gamma)$ and the constraint region for the coupling 
$\bar{\xi}^{D}_{N,\tau\tau}$. 

In future, with the reliable experimental result of 
$BR(\tau\rightarrow \mu\gamma)$ and obtaining more strict bounds for 
$BR(\mu\rightarrow e\gamma)$, $d_e$, $d_{\mu}$ and $d_{\tau}$ it would be 
possible to test models and free parameters underconsideration more
accurately. 
\section{Acknowledgement}
I would like to thank Prof. T. M. Aliev for useful discussions. 

\newpage
\newpage
\end{document}